\documentclass[epj]{svjour}
\usepackage{graphicx}
\usepackage{amsmath}
\usepackage{amssymb}
\usepackage{braket}
\usepackage{bigints}
\usepackage[table]{xcolor}

\begin{document}
\title{Split gluon masses in $SU(N)\times SU(M)$ theories}
\author{Julia G\'omez Concejo   \and Felipe J. Llanes-Estrada \and Diego Mar\'{i}a-Almaz\'an
\and Alexandre Salas-Bern\'ardez
}                     
\institute{Instituto de F\'{i}sica Te\'orica e IPARCOS, Facultad de CC. F\'{i}sicas, Plaza de las Ciencias 1, 28040 Madrid, Spain}
\date{\today}
%
\abstract{
We extend a known mass-gap equation for pure gluodynamics in global colour models (formulated in equal time quantization in Coulomb gauge) to one in which gluons split into two sets which may have different masses. 
If the theory is $SU(N)\times SU(M)$ with gluons in both groups having identical couplings (as suggested by Grand Unification arguments at large scales) it is immediate to see that different masses are generated for each subgroup. 
This global symmetry is not broken, but the split masses erase accidental symmetries that might be present due to the two couplings being the same at the large scale, such as $SU(N\times M)$ or similar.  
We also numerically explore a couple of low-dimensional examples of simple Lie groups, but in spite of the system 
having a form that would seem to allow spontaneous symmetry breaking, it is not triggered for these groups whose algebra has no ideal, and the dispersion relations for the various gluons converge to the same form.   
} 
\maketitle
\section{Introduction}  \label{intro}
Spontaneous gauge--boson mass generation is at the core of the Standard Model. Additionally to the Higgs mechanism, 
the Schwinger mechanism and similar ideas allow the gauge bosons (henceforth, ``gluons'') to acquire a mass without the assistance of an explicit additional field~\cite{Cornwall:1981zr,Cornwall:2015lna,Horak:2022aqx,Alkofer:2003jr,Fischer:2004ym}. 

Gluon masses are a welcome gauge--fixed feature of Chromodynamics as they raise glueballs from the low--lying hadron spectrum~\cite{Buisseret:2009yv,Vento:2004xx,Bugg:2000zy,Chen:2005mg} where the existing hadrons are well understood.

This is perhaps worth exploring in the context of Grand Unification because 
complicated symmetry breaking patterns~\cite{Lee:2016wiy,Mohapatra:1979nn} appear and the scalar Higgs-type mechanisms to break the symmetry can be convoluted (in fact, in many Grand Unification situations, the Higgs needs to be described by a composite field from the start~\cite{Dobson:2022crf,Dobson:2022ngz}).

We do not have a particular agenda nor unification model in mind, but want to generically explore a system of coupled gap equations that may allow splitting the gluon masses into two or more different values. Having this theoretical mechanism (which we partially achieve as will be explained in detail) would allow to have additional theoretical tools to explore unification dynamics. Because of the first theorem of Vafa and Witten~\cite{Vafa:1983tf}, we know that spontaneous global colour-symmetry breaking is impossible in the quark sector, so our exploration concentrates on the Yang-Mills sector alone.

Then there is also the question of why the Standard Model is built out of low-dimensional Lie groups~\cite{GarciaFernandez:2015jmn,Llanes-Estrada:2018azk} that may well have to do with the spontaneous acquisition of large masses (triggered by very different evolutions of the coupling constants) by particles charged under the (absent) large dimensional groups, which would remove such particles from the spectrum. 

These gap equations are formulated in Coulomb gauge, but our considerations should be easy to extend to other gauges such as Landau gauge~\cite{Cucchieri:2011ig,Li:2019hyv}. Modeling the Coul-omb gauge dynamics with simple global--colour model does not capture all the interesting phenomena, such as for example the Gribov divergent gluon mass at low momentum (a very strongly infrared enhanced propagator)~\cite{Reinhardt:2018dhg} but they are strong enough to trigger the generation of gluon masses: and we are not sure that we want to explore confinement (including Coulomb confinement is a necessary condition to describe confinement in arbitrary gauges, as the Coulomb potential is an upper bound for the QCD potential \cite{Zwanziger:2002sh}) in this work, that does not necessarily restrict itself to Quantum Chromodynamics (QCD) with the group $SU(3)$ but the production of a gap. This modified dispersion relation with a finite gluon mass is a feature of a more general class of theories.

In this article we review, in section~\ref{sec:simplegap}, the obtention of the known pure Yang-Mills gap equation in the North Carolina State~\cite{Szczepaniak:1995cw} model; we solve it for various groups, all of which have the same coupling constant at a low scale in section~\ref{sec:coloursymmetric}; we then, in section~\ref{section:rotura}, extend the mechanism to allow for the possibility of different variational wavefunctions for each of the gauge bosons, which could possibly trigger spontaneous breaking of a global symmetry. We succeed in doing this for product Lie groups or any other situation in which the underlying Lie algebra contains an ideal. Afterwards, we conduct a first numerical exploration for a few simple Lie groups of low--dimension, reported in section~\ref{sec:negativesimple} and do not currently find a situation in which the symmetry breaks.  After a brief outlook, we complement the discussion with an appendix detailing the numerical solution method, the necessary colour algebra, and an exhaustive list of the structure constant combinations (in the particular case of $SU(3)$ only).

\section{Coulomb--gauge gap equation for a singlet condensate} \label{sec:simplegap}
In this section we present the relatively well--known theory of the mass gap equation leading to a gluon mass in Coulomb gauge
with a color--singlet condensate (note that any gauge boson mass and condensates are necessarily features of a gauge--fixed picture of the theory) that therefore respects all global symmetries.
We start from a global--color symmetry preserving Hamiltonian~\cite{Szczepaniak:1995cw}:
\begin{align}
    H&=\int d^3\mathbf{x}\,(\mathbf{\Pi^a} \cdot \mathbf{\Pi^a} + \mathbf{B^a}\cdot \mathbf{B^a}) \nonumber- \\ &- \frac{1}{2}\int d^3\mathbf{x} \int d^3\mathbf{y} \rho^a_{\rm glue}(\textbf{x})V(|\textbf{x}-\textbf{y}|)\rho^a_{\rm glue}(\textbf{y})\ ,
    \label{eq:hamiltonian}
\end{align}
where $\mathbf{\Pi^a}$ represents the colour electric field, $\mathbf{B^a}$ the chromomagnetic field, and $\rho^a_{\rm glue}=f^{abc}\mathbf{A}^b\cdot\mathbf{\Pi}^c$ the colour charge density. The Hamiltonian differs from exact QCD in that the potential $V(|\textbf{x}-\textbf{y}|)$ 
is a c-function (like in electrodynamics) given below in Eq.~(\ref{eq:potencial}) simplifying the kernel that appears in the full non-Abelian theory~\cite{Christ:1980ku,Schwinger:1962wd}. 

In the basis of well-defined momentum particles, with creation, $a^{a\dagger}$, and destruction, $a^a$, boson operators the fields take the form
\begin{align}
    A^a_i(\mathbf{x})&=\int\frac{d^3\mathbf{k}}{(2\pi)^3}\frac{1}{\sqrt{2|\mathbf{k}|}}\left[a^a_i(\mathbf{k})+a^{a\dagger}_i(-\mathbf{k})\right]e^{i\mathbf{k\cdot x}}\label{eq:ABogoliubov}\\
    \Pi^a_i(\mathbf{x})&=-i\int\frac{d^3\mathbf{k}}{(2\pi)^3}\sqrt{\frac{|\mathbf{k}|}{2}}\left[a^a_i(\mathbf{k})-a^{a\dagger}_i(-\mathbf{k})\right]e^{i\mathbf{k\cdot x}}\label{eq:PiBogoliubov}\\
    B^a_i&=\epsilon_{ijk}\left(\bigtriangledown_jA^a_k+\frac{g}{2}f^{abc}A^b_jA^c_k\right)\;.
\end{align}
Because the Coulomb gauge is spatially transverse, adequate commutation rules that project out the longitudinal gluons are
\begin{equation}
    [a_i^a(\mathbf{k}),a_j^b(\mathbf{q)^\dagger}]=(2\pi)^3\delta^{ab}\delta^3(\mathbf{k}-\mathbf{q})\left(\delta_{ij}-\hat{k}_i\hat{k}_j\right)\; ,
\end{equation}
where $\hat{\mathbf{k}}\equiv \mathbf{k}/|\mathbf{k}|$.

A gluon condensed vacuum $\ket{\Omega}$ is variationally chosen by minimizing the expectation value of the Hamiltonian  $\langle H\rangle$. The quasiparticles that will annihilate it will have a dispersion relation $E(k)$ that serves as the actual variational function, controlling the canonical Bogoliubov rotation~\cite{Llanes-Estrada:2000cdq}
\begin{align}
    \alpha^a_i&=\cosh{\theta^a_k}a^a_i(\mathbf{k})+\sinh{\theta^a_k}a^{a\dagger}_i(\mathbf{-k})\\
    \alpha^{a\dagger}_i&=\sinh{\theta^a_k}a^a_i(\mathbf{k})+\cosh{\theta^a_k}a^{a\dagger}_i(\mathbf{-k})\;,
\end{align}
so that the field expansions become
\begin{align}
    A^a_i(\mathbf{x})&=\int\frac{d^3\mathbf{k}}{(2\pi)^3}\frac{1}{\sqrt{2E^a_{\mathbf{k}}}}\left(\alpha^a_i(\mathbf{k})+\alpha^{a\dagger}_i(-\mathbf{k})\right)e^{i\mathbf{k\cdot x}}\\
    \Pi^a_i(\mathbf{x})&=-i\int\frac{d^3\mathbf{k}}{(2\pi)^3}\sqrt{\frac{E^a_{\mathbf{k}}}{2}}\left(\alpha^a_i(\mathbf{k})-\alpha^{a\dagger}_i(-\mathbf{k})\right)e^{i\mathbf{k\cdot x}} \ .
\end{align}
The relation between the hyperbolic Bogoliubov angle  $\theta^a_k$ and the dispersion relation is then
\begin{equation}
    \tanh{\theta^a_\mathbf{k}}=\frac{|\mathbf{k}|-E^a_\mathbf{k}}{|\mathbf{k}|+E^a_\mathbf{k}} \ .
\end{equation}
Although it is not directly used in practice, the vacuum state of the interacting theory satisfying 
$\alpha_i^a\ket{\Omega}=0$ is obtained from the free vacuum via
\begin{eqnarray}
    \ket{\Omega}=e^{\left({-\int \frac{d^3\mathbf{k}}{2(2\pi)^3}}\tanh{\theta^a_\mathbf{k}}(\delta_{ij}-\hat{k}_i\hat{k}_j)a^{a\dagger}_i(\mathbf{k})a^{a\dagger}_j(-\mathbf{k})\right)}\ket{0}. \nonumber \\
\end{eqnarray}
To apply the Rayleigh-Ritz variational principle we require the  expectation value of the Hamiltonian in the family of rotated vacuum states,
\begin{align}
    \langle H_\Pi\rangle_\Omega&=(2\pi)^3\delta^3(0)\sum_a\int\frac{d^3\mathbf{k}}{(2\pi)^3}\frac{E^a_\mathbf{k}}{2}\nonumber \\
    \langle H_B\rangle_\Omega&=(2\pi)^3\delta^3(0)\sum_a\int\frac{d^3\mathbf{k}}{(2\pi)^3}\frac{|\mathbf{k}|^2}{2E^a_\mathbf{k}}\nonumber \\
    \langle H_V\rangle_\Omega&=(2\pi)^3\delta^3(0)\frac{1}{8}\sum_{abc}\int\frac{d^3\mathbf{k}}{(2\pi)^3}\frac{d^3\mathbf{k}'}{(2\pi)^3}
\nonumber\times \\\times&\left[(1+(\hat{\mathbf{k}}\cdot\hat{\mathbf{k'}}))\hat{V}(|\mathbf{k}-\mathbf{k'}|)f^{abc}f^{abc}\left(\frac{E^c_\mathbf{k}}{E^b_{\mathbf{k}'}}-C_G\right)\right] \ .
\end{align}
The factor $\delta^3(0)$ simply represents the quantization volume: it can be ignored in minimizing the energy density. The variational principle  then yields
\begin{equation*}
    \frac{\delta\langle H\rangle_\Omega}{\delta E^d_\mathbf{q}}=\frac{\delta(\langle H_\Pi\rangle_\Omega+\langle H_B\rangle_\Omega+\langle H_V\rangle_\Omega)}{\delta E^d_\mathbf{q}}=0
\end{equation*}
and this entails the following mass--gap equation for the gauge bosons,
\begin{align}
    (E^d_\mathbf{k})^2&=|\mathbf{q}|^2-\frac{1}{4}\sum_{a,b}f^{abd}f^{abd}\int\frac{d^3\mathbf{k}}{(2\pi)^3} \nonumber \times\\
&\times\hat{V}(|\mathbf{k}-\mathbf{q}|)(1+(\mathbf{\hat{k}} \cdot \mathbf{\hat{q}})^2)\left(\frac{(E^b_\mathbf{k})^2-(E^d_\mathbf{q})^2}{E^b_\mathbf{k}}\right).
    \label{eq:gapglue}
\end{align}
This is a nonlinear integral equation for $E_\mathbf{q}$ that appears also on the right, so the solution needs to be iterative until a fixed point is found. We can simplify a bit by noting that $E_\mathbf{k}$ does not depend on the angular variables, so that
defining an effective potential that absorbs the polar integral
\begin{equation}
    \hat{V}_{\text{eff}}(\mathbf{k},\mathbf{q})=\frac{1}{2\pi}\int d\Omega\,\hat{V}(|\mathbf{k}-\mathbf{q}|)(1+(\hat{\mathbf{k}}\cdot\hat{\mathbf{q}})^2),
\end{equation}
the radial equation becomes
\begin{align}
     (E^d_\mathbf{q})^2&=|\mathbf{q}|^2-\frac{1}{4}\sum_{a,b}f^{abd}f^{abd}\int_0^\infty\frac{d|\mathbf{k}|}{(2\pi)^2}|\mathbf{k}|^2 \nonumber \times\\
&\times\hat{V}_{\text{eff}}(\mathbf{k},\mathbf{q})\left(\frac{(E^b_\mathbf{k})^2-(E^d_\mathbf{q})^2}{E^b_\mathbf{k}}\right)\ .
    \label{eq:gapglueint}
\end{align}
An alternative way to derive this equation is via the functional approach~\cite{Watson:2013ghq}.

The potential $\hat{V}(|\mathbf{k}-\mathbf{q}|)$ is the Fourier transform of the potential $V(|\mathbf{x}-\mathbf{y}|)$ in Eq.~(\ref{eq:hamiltonian}). In a confining theory, the potential can be approximated by the Cornell linear+Coulomb $1/r$ potential, resulting in
\begin{equation}
    V(|\mathbf{x}-\mathbf{y}|)=-\frac{\alpha_s}{|\mathbf{x}-\mathbf{y}|}+b\,|\mathbf{x}-\mathbf{y}|\,e^{-\Lambda_{\rm phen}|\mathbf{x}-\mathbf{y}|}
    \label{eq:potencial}\;,
\end{equation}
where $\alpha_s$ is the strong interaction constant and $b$ the string tension. The term $e^{-\Lambda_{\rm phen}|\mathbf{x}-\mathbf{y}|}$ is a regulator that tames the strong infrared growth of the linear potential, but we will not use it and actually employ the computer grid to regulate the integration.

Since the gluon pairs in the condensate are in a singlet state, all quasiparticles have the same dispersion relation and no global symmetry is broken, $E^d_\mathbf{q}=E_\mathbf{q}$ for all $d$. The sum over the structure constants is the Casimir of the adjoint representation, $C_G=\sum_{a,b}f^{abd}f^{abd}$, that for $SU(N)$ is simply $C_G=N$. Eq.~(\ref{eq:gapglueint}) then reduces to
\begin{equation}
     (E_\mathbf{q})^2=|\mathbf{q}|^2-\frac{C_G}{4}\int_0^\infty\frac{d|\mathbf{k}|}{(2\pi)^2}|\mathbf{k}|^2\hat{V}_{\text{eff}}(\mathbf{k},\mathbf{q})\left(\frac{E_\mathbf{k}^2-E_\mathbf{q}^2}{E_\mathbf{k}}\right)\ .
    \label{eq:gapglueintnocolor}
\end{equation}

\section{Colour--symmetric gap equation for various groups} \label{sec:coloursymmetric}
Let us now separately study the effect of the terms of the potential in Eq.~(\ref{eq:potencial}), starting by the linear potential (first used in a gap equation, to our knowledge, in~\cite{Swift:1983fz,Adler:1984ri}),
\begin{equation*}
    V_L(|\mathbf{x}-\mathbf{y}|)=b\,|\mathbf{x}-\mathbf{y}|\;,
\end{equation*}
that, after Fourier transform, becomes
\begin{equation}
    \hat{V}_L=\mathcal{F}^3(V_L)=-\frac{8\pi b}{|\mathbf{k}-\mathbf{q}|^4}\;,
    \label{eq:Potencial Lineal Fourier}
\end{equation}
and handling the angular integrals yields the effective potential for the radial equation,
\begin{align}
    \hat{V}_{\text{eff},\,L}=&\int_{-1}^1d\theta \frac{-8\pi b}{(|\mathbf{k}|^2+|\mathbf{q}|^2-2|\mathbf{k}||\mathbf{q}|\cos{\theta})^2}(1+\cos^2\theta)
\nonumber= \\
   =&-8\pi b\Big[\left(\frac{|\mathbf{k}|^2+|\mathbf{q}|^2}{|\mathbf{k}|^2-|\mathbf{q}|^2}\right)^2\frac{1}{|\mathbf{k}|^2|\mathbf{q}|^2}+ \nonumber \\
  & +\frac{|\mathbf{k}|^2+|\mathbf{q}|^2}{4|\mathbf{k}|^3|\mathbf{q}|^3}\log\left(\frac{|\mathbf{k}|-|\mathbf{q}|}{|\mathbf{k}|+|\mathbf{q}|}\right)^2\Big]\;.
    \label{eq:potefflin}
\end{align}
The $k$ integral in Eq.~(\ref{eq:gapglueint}) has a log infrared divergence upon employing the  $1/(k-q)^4$. The regulated equation is numerically solved (as detailed in the appendix) and the solutions are plot in figure~\ref{fig:pot lineal} for different symmetry groups. 
In all cases we see the emergence of a mass, $m=E(0)$, larger with increasing dimension of the symmetry group due to the $C_G$ colour factor in Eq.~(\ref{eq:gapglueintnocolor}).

\begin{figure}[h]
\includegraphics[width=0.9\columnwidth]{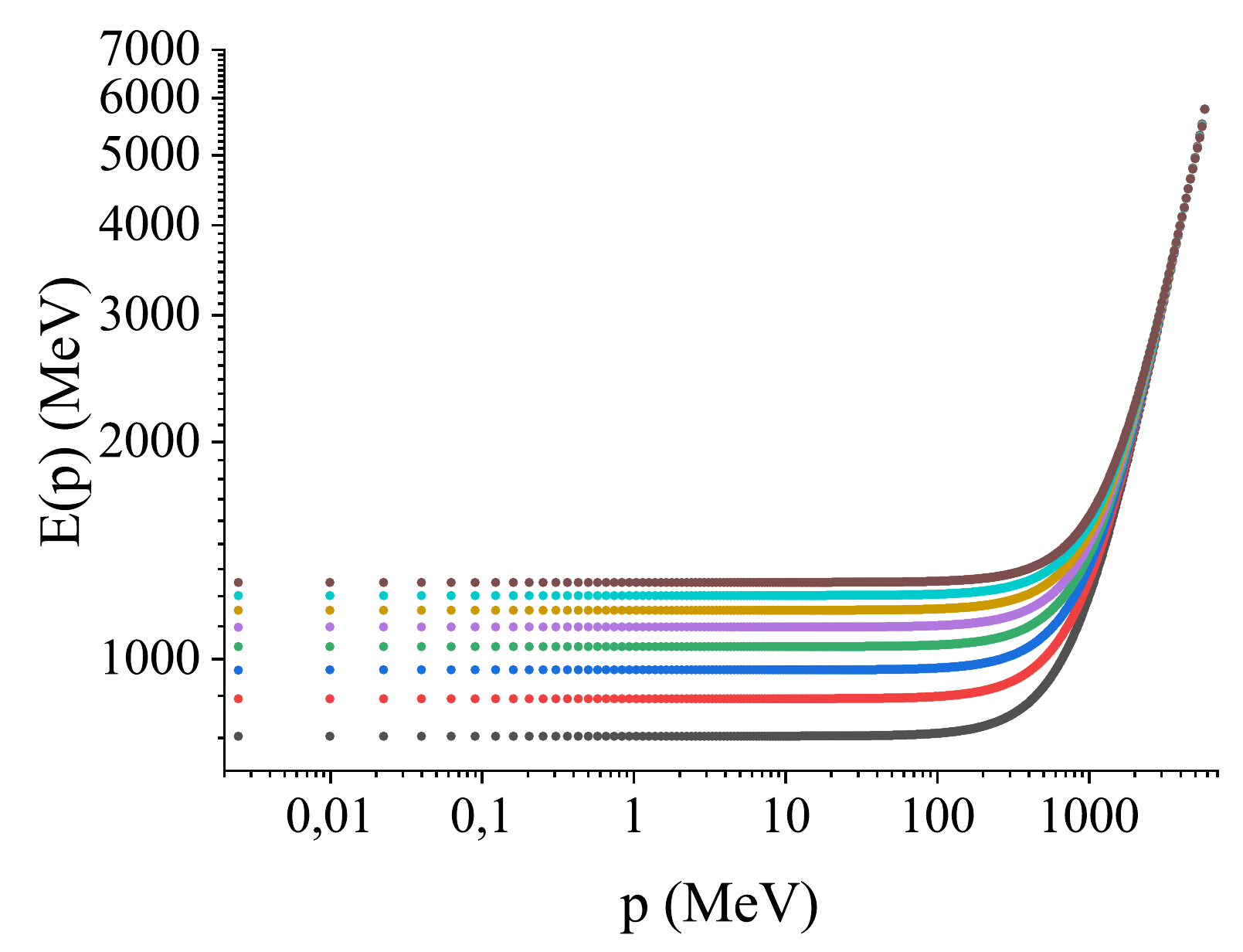}
    \caption{Computed numerical dispersion relations $E(p)$ with the linear potential from Eq.~(\ref{eq:potefflin}). 
From bottom to top they correspond to the groups $SU(3)$ through $SU(10)$, all with the same string tension $b=0.18$ GeV$^2$.
   \label{fig:pot lineal}}
\end{figure}

We now turn to the Coulomb potential that is a good description of the actual potential when interactions are small, that is, at high momentum transfers in non-Abelian theories.
It is 
\begin{equation*}
    V_C(|\mathbf{x}-\mathbf{y}|)=-\frac{\alpha_s}{|\mathbf{x}-\mathbf{y}|}\;,
\end{equation*}
with Fourier transform
\begin{equation}
    \hat{V}_C(|\mathbf{k}-\mathbf{q}|)=\mathcal{F}^3(V_C)=-\frac{4\pi\alpha_s}{|\mathbf{k}-\mathbf{q}|^2}\,,
\end{equation}
and, because of the absence of a $\phi$--dependence, just as for the linear potential, both being central, 
the effective potential for the radial equation is
\begin{align}\label{FTCoulomb}
    \hat{V}_{\text{eff},\, C}&=\int_{-1}^1d\theta \frac{-4\pi\alpha_s}{|\mathbf{k}|^2+|\mathbf{q}|^2-2|\mathbf{k}||\mathbf{q}|\cos{\theta}}(1+\cos^2\theta)=\nonumber\\
    =4\pi&\alpha_s\Big[\frac{1}{2|\mathbf{q}|^2}+\frac{1}{2|\mathbf{k}|^2}+\nonumber\\
    &+\frac{|\mathbf{k}|^4+6|\mathbf{k}|^2|\mathbf{q}|^2+|\mathbf{q}|^4}{8|\mathbf{k}|^3|\mathbf{q}|^3}\log\left(\frac{|\mathbf{k}|-|\mathbf{q}|}{|\mathbf{k}|+|\mathbf{q}|}\right)^2\Big]\ .
\end{align}
This potential causes no problem in the infrared $k\xrightarrow{}q$ limit, but the improper integral in the ultraviolet
 $k\xrightarrow{}\infty$ does not converge. Since, unlike the linear potential, the Coulombic one is scale--free,
the solutions scale with the regulating cutoff). Since it is not particularly appealing that the computer grid determines the mass gap (although common practice in many computer fields), we will eliminate that dependence by 
a fixed momentum subtraction  (MOM scheme). We therefore detract from Eq.~(\ref{eq:gapglueintnocolor}) the same equation but with a fixed value of the momentum scale, $\mu$, that is now dictating the solution's mass. The resulting equation, 
\begin{align} \label{gapsubtracted}
   & (E_\mathbf{q})^2=(E^d_\mu)^2+|\mathbf{q}|^2-\mu^2-\frac{C_G}{4}\int_0^\infty\frac{d|\mathbf{k}|}{(2\pi)^2}\frac{|\mathbf{k}|^2}{E_\mathbf{k}}\times
\nonumber \\ \times&(\hat{V}_{\text{eff}}(\mathbf{k},\mathbf{q})\big((E_\mathbf{k})^2-(E_\mathbf{q})^2\big)-  \hat{V}_{\text{eff}}(\mathbf{k},\mu)\big((E_\mathbf{k})^2-(E_\mu)^2\big)\ , \nonumber \\
\end{align}
has a much better behaviour, and any $k\xrightarrow{}\infty$  integration divergence is suppressed by the new fixed-point subtraction, with the same potential but opposite sign. $\mu$ has to be chosen high enough so that the energy 
be practically equal to the momentum, that is, $E(\mu)=\mu$, and one is in a quasifree regime. Beyond that, $\mu$ is arbitrary
just as the choice of cutoff was. Still it allows control of the problem's scale without regards to the integration grid.

\begin{figure}[h]
    \centering
\includegraphics[width=\linewidth]{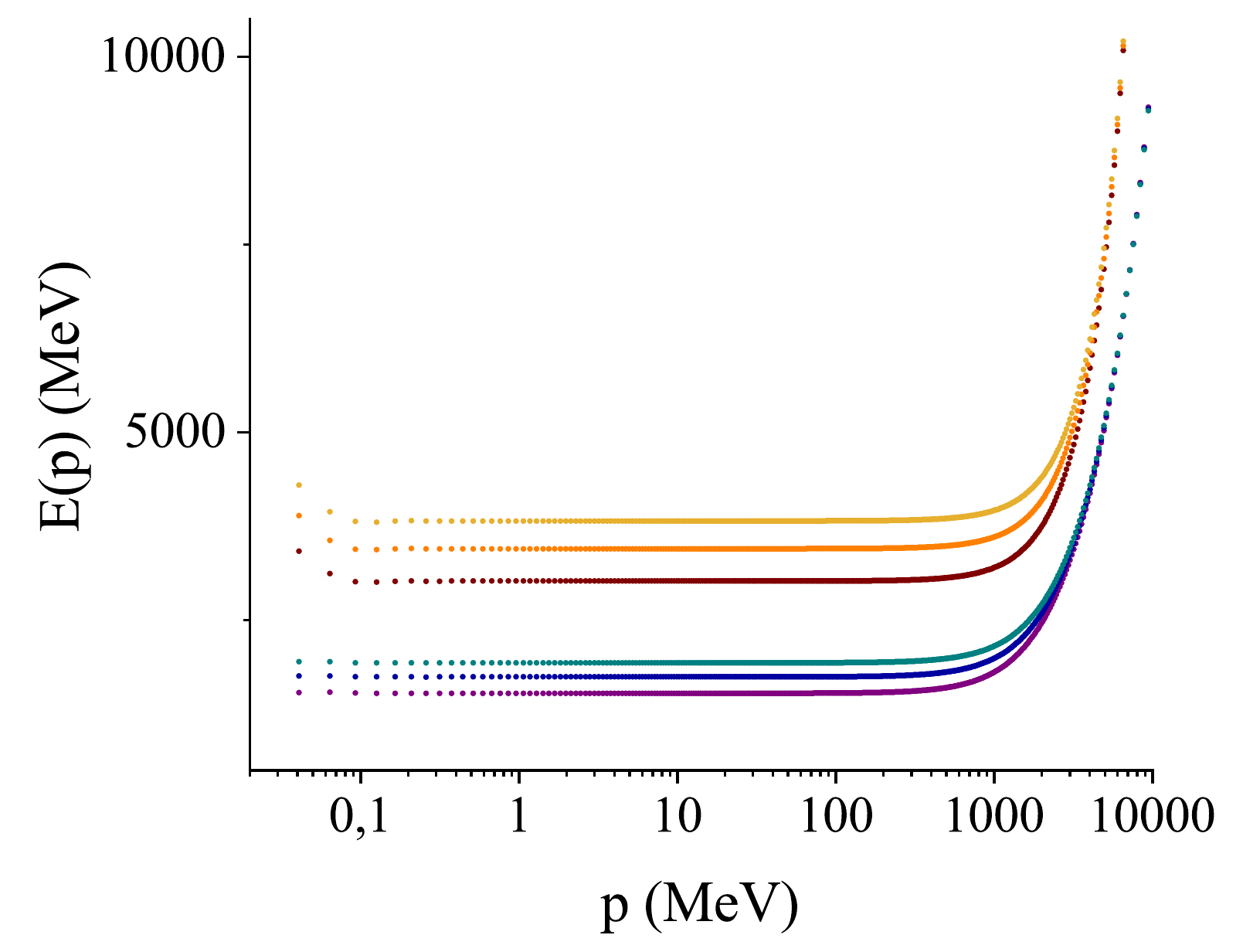}
\includegraphics[width=\linewidth]{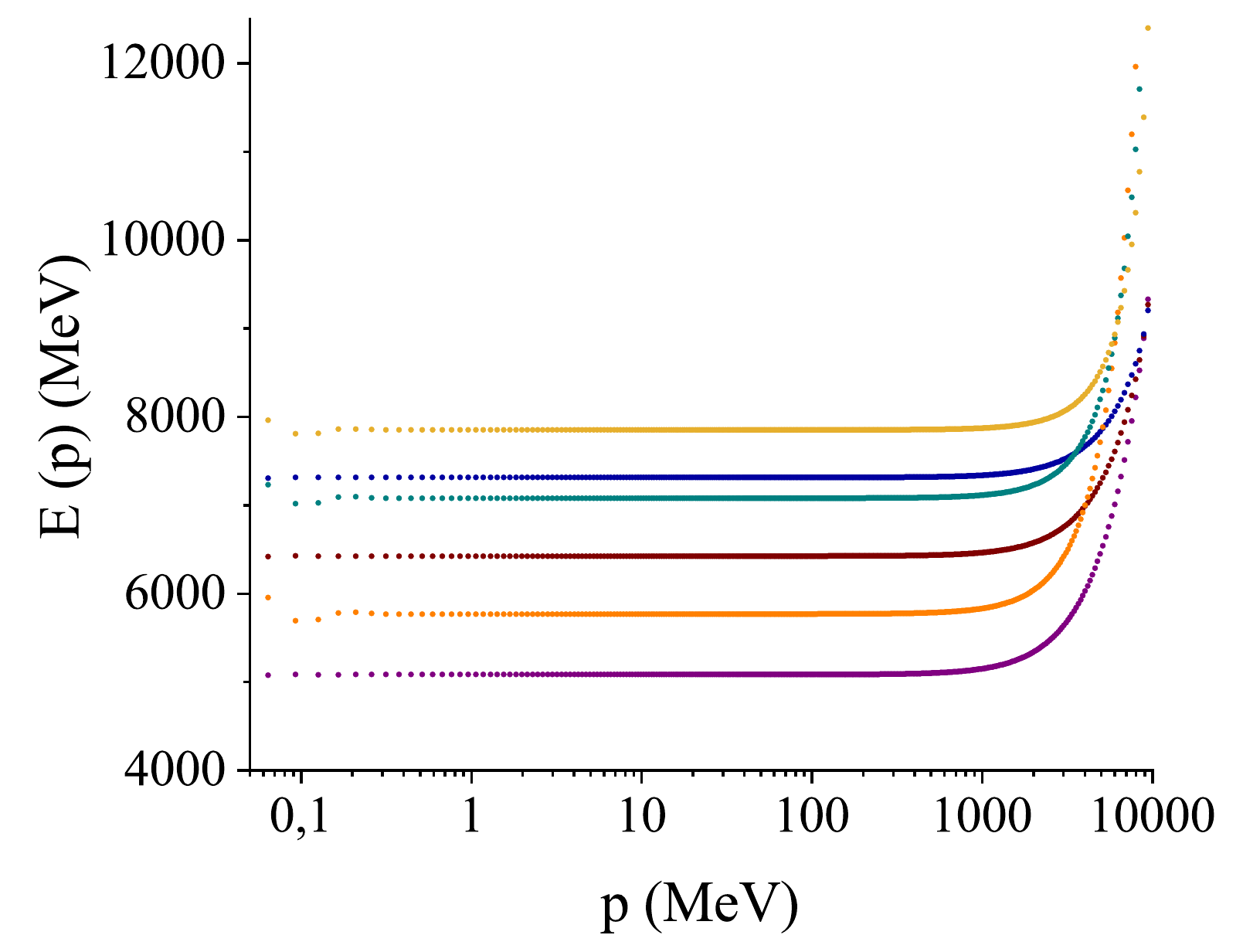}
 \caption{Dispersion relations solving the gauge--boson gap equation~(\ref{eq:gapglueintnocolor}) 
for different $SU(N)$, all with the same strong coupling constant $\alpha_s=1$ in Eq.~(\ref{FTCoulomb}), with cutoffs $k_1=10$ GeV and $k_2=20$ GeV.
The top plot (regularized but unrenormalized equation) shows two bunches of three functions $E(k)$, with the lower bunch employing the regulator $k_1$ and the upper one using $k_2$; in both cases the symmetry groups correspond to $SU(3)$, $SU(4)$ and $SU(5)$. The bottom plot shows the same solutions but now with a MOM subtraction and renormalization scale of $\mu=9$ GeV.}
   \label{fig:renorm}
\end{figure}

Upon applying this method to various symmetry groups, the masses generically diminish, as can be observed in figure~\ref{fig:renorm}.  The qualitative features of mass generation are similar to the linear potential, and 
in both cases the larger the dimension of the symmetry group, the larger the mass which is generated, all other things being equal, due to the $C_G$ colour factor. 

Of course, the scales of both plots have been set so that the resulting gluon masses make sense at the QCD scale with $SU(3)$, yielding glueballs of reasonable mass \cite{Llanes-Estrada:2000ozq}, but the reader can easily scale them as needed to any other energy regime. 
The resulting dispersion relations are standard, as would appear in a plasma with a cutoff or upon solving the Helmholtz equation in a waveguide, and show that a minimum threshold energy is required to propagate gluons of any momentum.

The gluon masses generated for large--dimensional groups are not exponentially far from the QCD one, but this is because
we start with the same coupling constant at a low scale. If instead we started with the same coupling constant at 
a very large (Grand Unification) scale, the much larger antiscreening of the Yang-Mills coupling constant for larger groups
would yield exponentially larger masses at a low--scale, effectively removing such theories from the spectrum, as we have shown elsewhere~\cite{GarciaFernandez:2015jmn,Llanes-Estrada:2018azk}.

\section{Splitting the masses in $SU(N)\!\!\times\!\! SU(M)$}\label{section:rotura}

The mass generation that we have so far fixed forced us to fix the local gauge (we have adopted the Coulomb one, but 
similar results have been obtained in others), but the solutions fully respect the global color symmetry.
In this section we turn to the possibility that the solutions may spontaneously break some global symmetry 
and different gluons come with different masses even if the scale $\mu$ and the coupling $\alpha_s$ are the same for all of them.

We will, for the sake of simplicity, explore the partition of the $N^2-1$ gluons of $SU(N)$ in two subsets, one 
with  $n$ lighter gluons and another, containing the rest of them, heavier.
We need a bit more notation to distinguish the sets, and have opted for lowercase letters
 $a,b,c,d...=1,...,n$ to denote the colours of the lighter ($L$) gluons, whose dispersion relation shall be written as
 $\omega_\mathbf{q}$, and uppercase letters  $A,B,C,D...=n,...,N^2-1$ for the heavier ($H$) ones, with dispersion relation naturally
chosen as the capital letter $\Omega_\mathbf{q}$. To refer to all the colors simultaneously we adopt the greek indices
 $\alpha,\beta,\gamma,\delta...=1,...N^2-1$.

The gap equation~(\ref{eq:gapglueint}) now formally separates into a $2\times 2$ system for the two types of gluons,
\begin{align}
   & (\omega_\mathbf{q}^d)^2=|\mathbf{q}|^2-\frac{1}{4}\int_0^\infty\frac{d|\mathbf{k}|}{(2\pi)^2}|\mathbf{k}|^2\hat{V}_{\text{eff}}(\mathbf{k},\mathbf{q}) \sum_\alpha \nonumber \\ 
&\left[\sum_{b}(f^{\alpha bd})^2\frac{(\omega^b_\mathbf{k})^2-(\omega^d_\mathbf{q})^2}{\omega^b_\mathbf{k}}+\sum_{B}(f^{\alpha Bd})^2\frac{(\Omega^B_\mathbf{k})^2-(\omega^d_\mathbf{q})^2}{\Omega^B_\mathbf{k}}\right] \nonumber \\ \\ \nonumber 
   & (\Omega_\mathbf{q}^D)^2=|\mathbf{q}|^2-\frac{1}{4}\int_0^\infty\frac{d|\mathbf{k}|}{(2\pi)^2}|\mathbf{k}|^2\hat{V}_{\text{eff}}(\mathbf{k},\mathbf{q}) \sum_\alpha\\
&\left[\!\sum_{B}\!(f^{\alpha BD})^2\!\frac{(\Omega^B_\mathbf{k})^2\!-\!(\Omega^D_\mathbf{q})^2}{\Omega^B_\mathbf{k}}\!+\!\sum_{b}\!(f^{\alpha bD})^2\!\frac{(\omega^b_\mathbf{k})^2\!-\!(\Omega^D_\mathbf{q})^2}{\omega^b_\mathbf{k}}\!\right] \nonumber \;.\\
    \label{eq:SistemaRoturaForzada}
\end{align}

Each integral contains two terms between brackets. The first depends only on the dispersion relation being solved 
for on the left hand side of the equation (diagonal terms), be it  $\omega$ for the light bosons or  $\Omega$ for the heavy ones. The second term depends on both dispersion relations and couples the equations, pushing the solution
towards the symmetric point $\omega^a=\Omega^A$ (this can be seen with a little patience from the combination of signs).

More compactly, and focusing on the colour structure of these equations, this system reads
\begin{align}\label{eq:sistemapp}
		(\omega_\mathbf{q}^d)^2&=|\mathbf{q}|^2-\frac{1}{4}\bigintsss\frac{d|{\mathbf{k}|}}{\left(2\pi \right) ^2}|\mathbf{k}|^2
\hat{V}_{\text{eff}}(\mathbf{k},\mathbf{q}) \times\nonumber \\
&\times\left(\text{LL}\sum_{\alpha,b}\left( f^{\alpha bd}\right) ^2+\text{LH}\sum_{\alpha,B}\left( f^{\alpha Bd}\right) ^2\right)
\nonumber \;,\\ 
		(\Omega_\mathbf{q}^D)^2&=|\mathbf{q}|^2-\frac{1}{4}\bigintsss\frac{d|{\mathbf{k}|}}{\left(2\pi \right) ^2}|\mathbf{k}|^2 \nonumber \times\\
&\times\left(\text{HL}\sum_{\alpha,b}\left( f^{\alpha bD}\right)^2+\text{HH}\sum_{\alpha,B}\left( f^{\alpha BD}\right) ^2\right)
	 . \nonumber \\
\end{align}
where the various symbols have obvious meaning shortening Eq.~(\ref{eq:SistemaRoturaForzada}).

The observation that gives this paper its title is that, if the combination of structure constants of these coupling terms
would vanish, which can be achieved by making all constants of the forms $f^{\alpha Bd}$ and $f^{\alpha bD}$
to vanish, the two equations completely decouple. 
We then have two copies of Eq.~(\ref{eq:gapglueint}), one for the light dispersion relation $\omega(q)$ and one for the heavier $\Omega(q)$. 
The mass in each case is proportional to the size of the factors
$\sum_{\alpha,b}(f^{\alpha bd})^2$ or $\sum_{\alpha,B}(f^{\alpha BD})^2$ that appear in the diagonal terms 
(not all four terms can simultaneously vanish in a non-Abelian gauge theory, since they are sums of squares and some
structure constants must be nonzero).

The vanishing of the two coupling terms is precisely what happens if the split of the gauge bosons is done along the lines
of two commuting generating algebras, so that the algebra corresponding to the total group is not simple, but the
direct sum of two ideals, in group--theory parlance ($\mathfrak{su}(N)\oplus\mathfrak{su}(M)\oplus...$).

We have chosen to split the system in two sets, but the reader can easily note that a larger number of dispersion relations
$\omega_1$, $\omega_2$,\dots $\omega_j$ is possible, in which case the system of equations would further split into several. For each ideal in which we can decompose the algebra, we will obtain one decoupled equation that will provide a different gauge--boson mass, as long as the dimensions are different,  $N\neq M$, which drive different colour factors.

This splitting happens even in the presence of the same effective potential, coupling constant and renormalization scale, and is entirely driven by the colour factors.

In the case of simple Lie algebras such as $\mathfrak{su}(N)$, without proper ideals, this decoupling cannot take place (because no subset of generators of the algebra can commute with all of those outside the subset, in which case some of the mixed--index $f$ structure constants need to be different from zero), and the various gap equations are necessarily coupled to one another. To this we turn in the next section.

\section{Global colour breaking not obvious for a simple Lie algebra} \label{sec:negativesimple}
In this section we report an initial exploration of the new system of coupled equations for a couple of low-dimensional Lie algebras. Because there is no ideal, at least three of the four terms  contribute, independently of how the partition of the $N^2-1$  gluons is taken. Thus, the system remains coupled.

\begin{figure}[ht]
\centering
        \includegraphics[width=0.9\columnwidth]{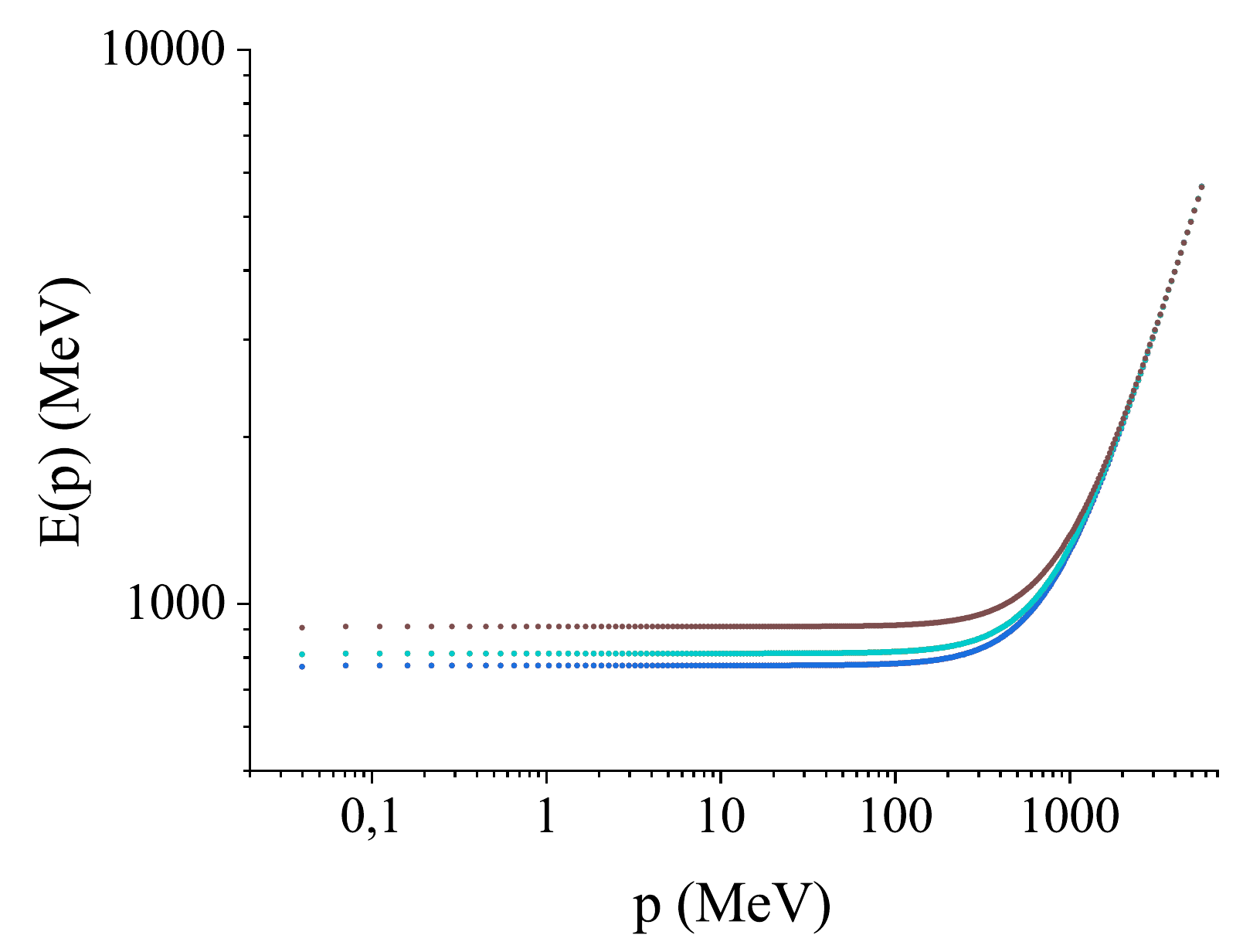}
        \includegraphics[width=0.9\columnwidth]{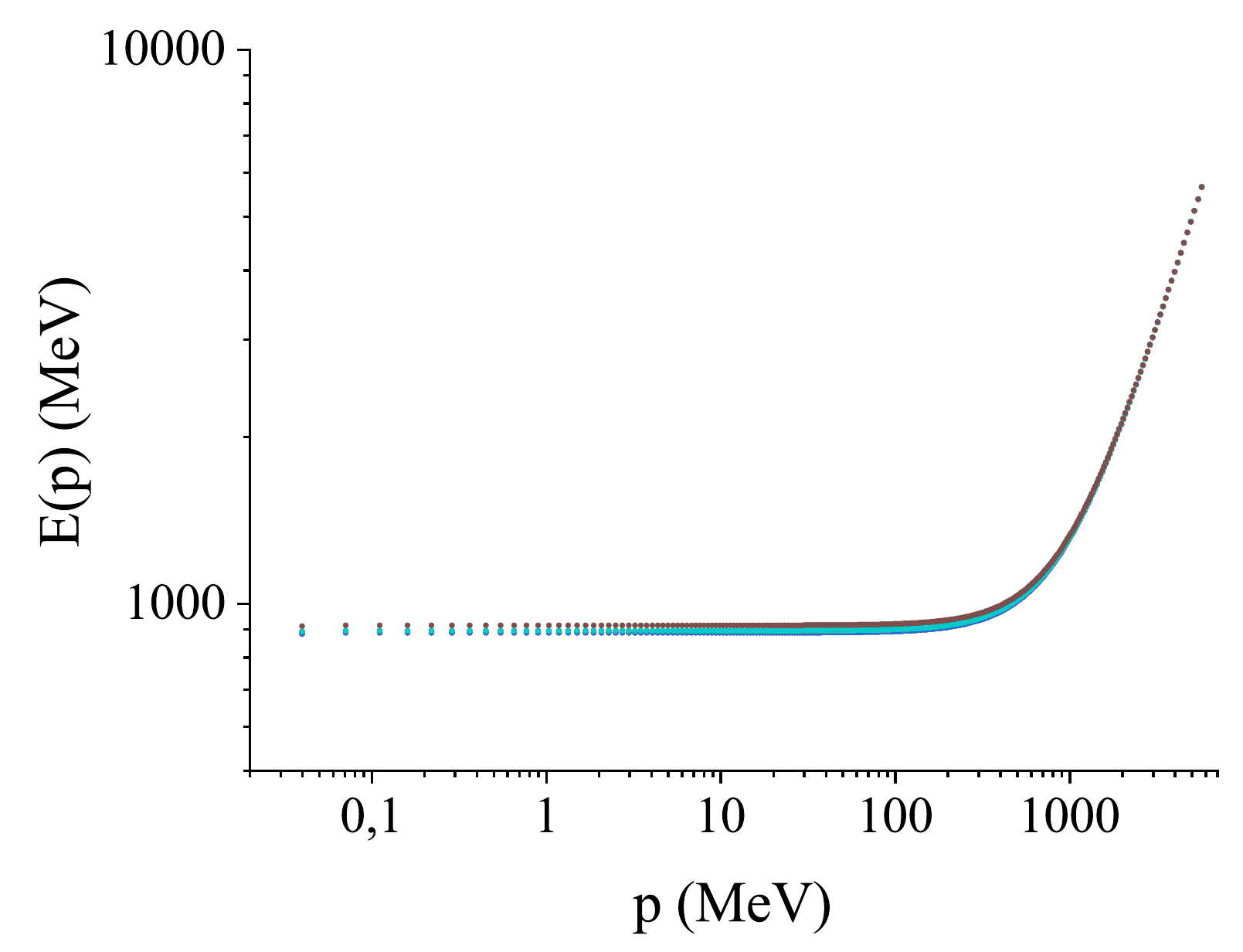}
        \caption{We artificially attenuate the nondiagonal coupling among dispersion relations in Eq.~(\ref{eq:SistemaRoturaForzada}) to show a reaction of the system breaking global colour. Upper plot: the multiplicative
factor is 0.1, greatly damping the strength of the coupling, and thus showing enhanced symmetry breaking. 
Lower plot: the factor is only 0.8, and the explicit symmetry breaking is still visible. The symmetry breaking pattern is $SU(3)\to SU(2)$.}
    \label{fig:rotura01}
\end{figure}

In a first exercise, we attempt to break the symmetry by hand. In a totally artificial manner, we include a multiplicative factor in the off-diagonal terms of Eq.~(\ref{eq:SistemaRoturaForzada}) that reduces their intensity.
The outcome is exposed in figure~\ref{fig:rotura01}, both for a strong artificial suppression by 1/10 but also for a modest 
reduction factor of 8/10. In both cases, the global colour symmetry is broken and the system converges to two dispersion relations with different mass. The plots correspond to a global $SU(3)$ colour group in which an $SU(2)$ subgroup 
remain light (dispersion relation $\omega$) and the rest acquire a heavier dispersion relation $\Omega$. 

However, if we reset the equation to its original form without artificial factors explicitly breaking the symmetry, 
it is not so easy to find a solution with spontaneous breaking of a simple group. 
We have not yet deployed this project to a supercomputer; in a tabletop machine we have been able to quickly examine the following symmetry breaking chains: $SU(2)\to U(1)$, $SU(3)\to SU(2)$, $SU(4)\to SU(3)$, $SU(4)\to SU(2)$ and $SU(5)\to SU(4)$. For example, in this last case, of the $5^2-1=24$ bosons, $4^2-1=15$ were candidates to remain light and the remaining 9 candidates to become heavy. 

As an example analysis, we provide detail for a partition of the 8 gluons of $SU(3)$ into a group of 3 and a group of 5. Depending on how the first group is chosen, its three gluons may correspond to a subgroup $SU(2)$.

Table~\ref{tab:fconstants1} in the appendix lists the sums of squared structure constants of the nondiagonal, coupling terms, 
those multiplying $LH$ and $HL$ in Eq.~(\ref{eq:sistemapp}) for  possible combinations of three gluons chosen among the eight of $SU(3)$. 

\begin{figure}[h]
	\centering  
		\includegraphics[width=0.9\columnwidth]{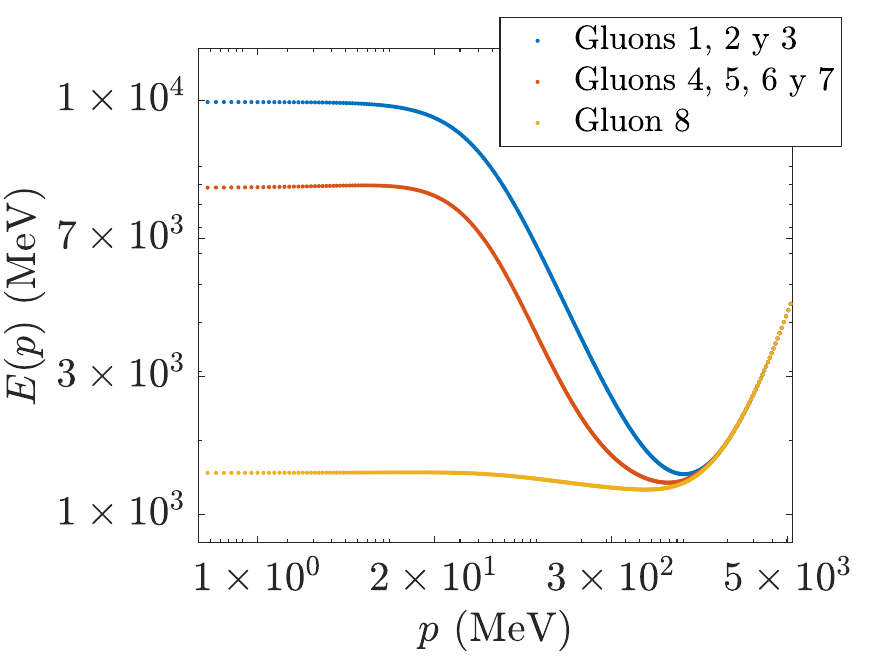}
		\includegraphics[width=0.9\columnwidth]{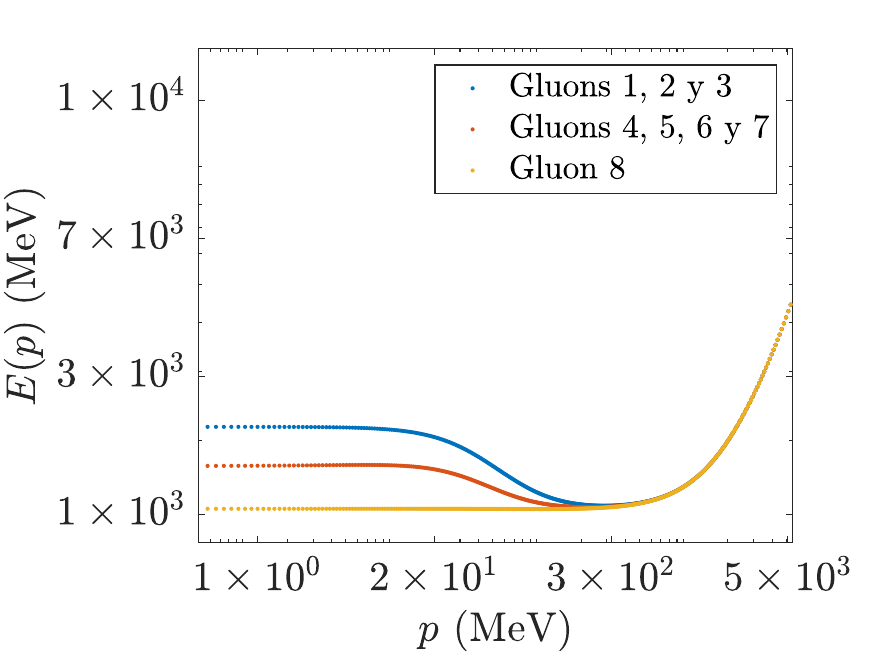}
		\includegraphics[width=0.9\columnwidth]{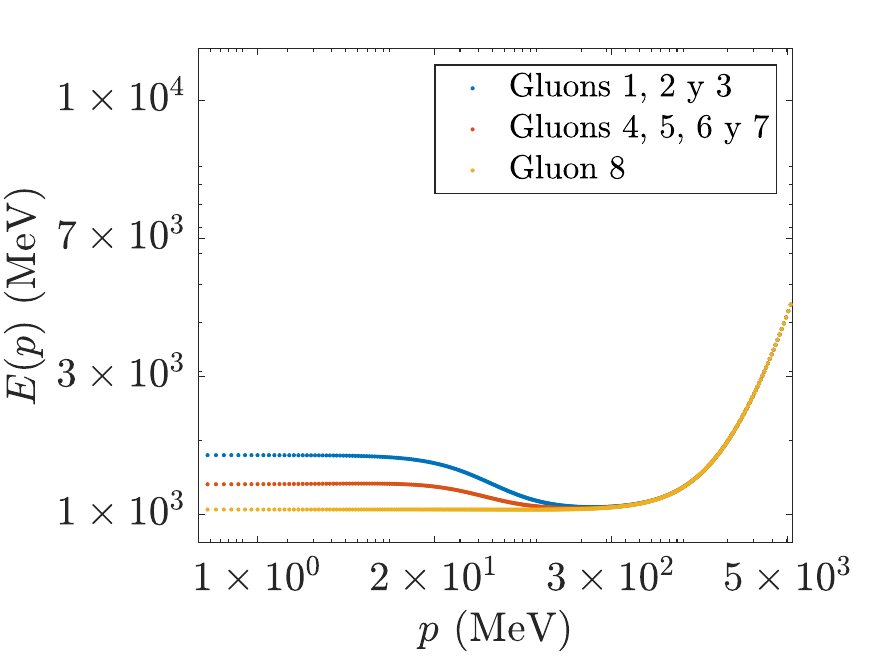}
	\caption{Intermediate steps towards convergence for the dispersion relations of $SU(3)$ gluons
with initially disparate masses.
Shown are calculations at 1, at 100 and at 200 iterations.
The gluon numbered as 8 is seen to quickly decouple from the rest even on the upper diagram, the first iteration. 
With 100, it is clear that the system is already converging towards the symmetric solution.
	\label{fig:desmarcodel8}}
\end{figure}

As an example, reading the first row of the table, we observe that the coupling of gluon number 8 is null. This means that, initially, this gluon's dispersion relation does not converge towards the others. However, the rest of the system is 
coupled to it in a nonvanishing way (because the corresponding generator $T^8$ is not nor does it belong to an ideal of the underlying algebra), so that the system evolves towards the symmetric solution. This can be observed in figure~\ref{fig:desmarcodel8}.
 
To start the iteration we have chosen an initial mass of  
42.4 MeV for the first three gluons and 848 MeV for the rest (the precise choice
of these numbers is immaterial, they here have to do with the units employed in the program, 424 MeV as the
scale of the string tension when the linear potential is active, not the case in this purely Coulombic computation).
After initial large jumps, that even change the second derivative of the dispersion relations in intermediate steps, the gluons converge towards the same mass in the 1 GeV range, in this case.

One could then conjecture that the system of Eq.~(\ref{eq:SistemaRoturaForzada}) has as only fixed point $\omega^a=\Omega^A=\omega$ for all gluons, due to some symmetry upon reorganizing the $f^{abc}$ structure constants among different choices of the boson partition. Whatever this might be, we have not yet been able to identify it. 

\newpage
\section{Outlook}

We have worked out the known gap equation for gluodynamics in the North Carolina State family of global colour models inspired by Coulomb gauge QCD, then extended it to allow for the possibility of different bosons acquiring different mass 
in a spontaneous way. This leads to a coupled system of equations and we have performed a first investigation of its colour structure. For simplicity, we have limited ourselves here to split the gluons into two groups (light and heavy),
but we have also made a few exploratory runs in which each of the gluons might acquire its own different mass. Computations here are more numerically costly as several independent functions have to be simultaneously determined (and note that 
the number of them grows as $N^2-1$ with the dimension of the group). We have found nothing different to report so we omit the discussion for the time being. 

What would be extremely interesting is to find an alley for spontaneous symmetry breaking among simple groups such as $SU(N)\to SU(M)$ but we have not yet identified an example where the system converges to two sets of gluons with different mass in this form. If this was possible, one could do away with complicated Higgs boson representations in Grand Unified Theories that seem rather ad hoc. 
Our exploratory study has been limited in scope and we have only examined a few group breaking patterns. 

Because we do not have a clear proof that the system must remain symmetric for a simple group either, we must limit ourselves to leaving the question of whether this is possible as open for future investigation. 
With more computing power we hope to be able to systematize the choice of the gluons that remain vs. those that remain heavy,
to extend the system beyond binary (with three, four or more types of dispersion relations with different gluon masses)

The current findings do show how, due to the colour factors alone (with equal couplings for the two groups),
the gauge bosons in an $SU(N)\times SU(M)$ theory with $N\neq M$ acquire different fixed--gauge masses.
This entails a breaking of possible accidental symmetries:
for example, fermions in the fundamental representation carrying indices for both groups
$\psi_{nm}$ could be seen, stretching the index, (such as in 1=red--up, 2=red--down, 3=blue--up etc.)
as belonging to the fundamental representation of $SU(N\times M)$. This global symmetry is not gauged by definition of the Lagrangian, rather it would be ``accidental''. It ceases making sense when the mass of the gauge bosons
of the two subgroups are different, so it is broken without the resort of a Higgs boson multiplet. 
As argued by Dobson and collaborators~\cite{Dobson:2022crf,Dobson:2022ngz}, the Higgs in more general theories than the SM should best be defined in terms of composite, gauge--invariant fields. It is not unconceivable that in strongly coupled theories, the equivalent field could be made from modes in the gauge--boson spectrum itself, due to nonlinearities.

\clearpage

\section*{Appendix}
\subsection*{Numerical solution of the gap equation}

In this first appendix we comment on the numerical solution of Eq.~(\ref{eq:gapglueint}). The method is easily extended to the system~(\ref{eq:sistemapp}).

We proceed by iteration from an initial guess $\Tilde{E}^d(q)$ for the dispersion relation that differs from the 
real function by  $E^d(q)=\Tilde{E^d}(q)+\epsilon^d(q)$ (where we define $q\equiv |\mathbf{q}|$). This we substitute in Eq.~(\ref{gapsubtracted})  
to isolate $\epsilon^b(q)$ to linear order in the Taylor expansion (we here omit the renormalization subtraction for 
conciseness, but it has been programmed),
\begin{align}
    \Tilde{E}^d(q)^2-q^2+\frac{1}{4}\sum_{a,b}f^{abd}f^{abd}
\int_0^\infty\frac{dk}{(2\pi)^2}k^2\hat{V}_{\rm eff}(k,q)
\times \nonumber \\ 
\times\left(\frac{\Tilde{E}^b(k)^2-\Tilde{E}^d(q)^2}{\Tilde{E}^b(k)}\right)\approx \nonumber \\
-2\Tilde{E}^d(q)\epsilon^d(q)-\frac{1}{4}\sum_{a,b}f^{abd}f^{abd}  \int_0^\infty\frac{dk}{(2\pi)^2}k^2\hat{V}_{\rm eff}(k,q)  \nonumber \times \\
\times\left[\left(\frac{\Tilde{E}^b(k)^2+\Tilde{E}^d(q)^2}{\Tilde{E}^b(k)^2}\epsilon^b(k)\right)-2\frac{\Tilde{E}^d(q)\epsilon^d(q)}{\Tilde{E}^b(k)}\right]\ . \nonumber \\
    \label{eq:numericGapglue}
\end{align}
It is convenient to introduce auxiliary functions,
    \begin{align}
        b^d(q)&:=\Tilde{E}^d(q)^2-q^2+\frac{1}{4}\sum_{a,b}f^{abd}f^{abd}\times \nonumber \\
&\times\int_0^\infty\frac{dk}{(2\pi)^2}k^2\hat{V}_{\rm eff}(k,q)\left(\frac{\Tilde{E}^b(k)^2-\Tilde{E}^d(q)^2}{\Tilde{E}^b(k)}\right)
\nonumber \\
A^{db}(q,k)&:=-2\Tilde{E}^b(k)\delta^{bd}\delta(k-q)-\nonumber \\
&-\frac{1}{4}\sum_{a}f^{abd}f^{abd}\frac{k^2}{(2\pi)^2}\hat{V}_{\rm eff}(k,q)\times
\nonumber \\
&\times\left[\left(\frac{\Tilde{E}^b(k)^2+\Tilde{E}^d(q)^2}{\Tilde{E}^b(k)^2}\right)-2\frac{\Tilde{E}^d(q)\delta^{bd}\delta(k-q)}{\Tilde{E}^b(k)}\right]   \label{eq:numeric gluons}\;,
    \end{align}
  to shorten notation, so that Eq.~(\ref{eq:numericGapglue}) is recognisable as a linear system
\begin{equation}
    b^d(q)=\sum_b\int_0^\infty dk\,A^{db}(q,k)\epsilon^b(k)\ .
\end{equation}

We then discretise momenta to make the expression amenable to automation,
\begin{equation}\label{solvenondiagonal}
    b^d_i=\sum_b\sum_j \Delta k_j A^{db}_{ij}\epsilon^b_j \ .
\end{equation}
As $E(k)$ approaches its linear asymptote for large $k$ and its nontrivial structure is at low $k$, 
we skew the discrete grid to have more points towards low $k$ with the help of a change of variable that introduces a Jacobian. 

Here the algorithms for Eq.~(\ref{eq:gapglueint}) and (\ref{eq:sistemapp}) diverge, as the global colour symmetry (all gluons have equal mass) allows the first to take a simpler form. Because all $\omega^a$ are equal, they can be factored out of the colour factors, so that the colour indices can be summed with the closure relation. The auxiliary quantities of Eq.~(\ref{eq:numeric gluons}) then read
 \begin{align}   
        b(q)&:=b^d(q)=\Tilde{E}(q)^2-q^2+\frac{C_G}{4}\int_0^\infty\frac{dk}{(2\pi)^2}k^2 \times\nonumber\\ &\times\hat{V}_{\rm eff}(k,q)\left(\frac{\Tilde{E}(k)^2-\Tilde{E}(q)^2}{\Tilde{E}(k)}\right)\nonumber\\
        A(q,k)&:=\sum_bA^{db}(q,k)=-2\Tilde{E}(k)\delta(k-q)-\frac{C_G}{4}\frac{k^2}{(2\pi)^2}\times \nonumber\\&\times \hat{V}_{\rm eff}(k,q)\left[\left(\frac{\Tilde{E}(k)^2+\Tilde{E}(q)^2}{\Tilde{E}(k)^2}\right)-2\frac{\Tilde{E}(q)\delta(k-q)}{\Tilde{E}(k)}\right] \ .
 \end{align}
The linear system that allows to extract each Newton step of the algorithm then takes the form
\begin{equation}
    b(q)=\int_0^\infty dk\,A(q,k)\epsilon(k)\;,
\end{equation}
or discretized,
\begin{equation}
    b_i=\sum_j \Delta k_j A_{ij}\epsilon_j\, ,
\end{equation}
which is solved for $\epsilon$, allowing for the update of $E(k)$.
In the case of nondiagonal colour couplings, Eq.~(\ref{solvenondiagonal}) has to be addressed instead.

\subsection*{Structure constants for the generic Lie algebra $\mathfrak{su}$(N)}\label{section:numericStr}
The $SU(N)$ has $N^2-1$ generators that commute according to the rules of the  $\mathfrak{su}$(N) algebra, 
$[T^a,T^b]=if^{abc}T^c$ with structure constants  $f^{abc}$.

The following formulae give these structure constants in a direct way which makes them apt for computer programming,
as necessary to solve for example Eq.~(\ref{eq:gapglue}).

The $N^2-1$ generators of  $SU(N)$ can be split into three subsets. First, there are the $N-1$ diagonal matrices of the Cartan subalgebra, that commute and thus provide simultaneous good quantum numbers (such as hypercharge and the third component of isospin in the case of $SU(3)$). Then, there are~\cite{Bertlmann_2008}  $N(N-1)/2$ antisymmetric matrices and $N(N-1)/2$ symmetric but nondiagonal matrices. We then split the color index of the adjoint representation into three distinct ones for each of this types of matrices: $D$ for diagonal, $S$ for symmetric and $A$ for antisymmetric. When acting on the fundamental representation of the fermions they take the explicit form~\cite{Bertlmann_2008}:
\begin{align}
    T_{S_{nm}}&=\frac{1}{2}(\ket{m}\bra{n}+\ket{n}\bra{m})\\
    T_{A_{nm}}&=\frac{1}{2}(\ket{m}\bra{n}-\ket{n}\bra{m})\\
    T_{D_n}&=\frac{1}{\sqrt{2n(n-1)}}\left(\sum_{k=1}^{N-1}\ket{k}\bra{k}+(1-n)\ket{n}\bra{n}\right)\;.
\end{align}

 The $n$ and $m$ subindices take $N$ different values, the size of the fundamental representation of  $\mathfrak{su}$(N).
We can retrieve the values of the various subindices from the closed formulae
\begin{align*}
    S_{nm}&=n^2+2(m-n)-1\\
    A_{nm}&=n^2+2(m-n)\\
    D_n&=n^2-1\;,
\end{align*}
that guarantee that no value is repeated and all values from  $1$ to $N^2-1$ is covered when additionally imposing $1\leq m <n\leq N$. Thus, any  $a\in\{S_{nm},A_{nm},D_{n}\}$ and the correspondence between these and the usual indices is bijective. 

For example, in $SU(2)$ one has three generators, and with this convention the symmetric one is $T_{S_{21}}=T_1$ (Pauli's $\sigma_x/2$ matrix, essentially), the antisymmetric $T_{A_{21}}=T_2$ (that is, $\sigma_y/2$) and the diagonal one is $T_{D_2}=T_3$ (or $\sigma_z/2$).
If, in turn, we now apply these indexing rules to $SU(3)$, we reproduce Gell-Mann's matrices in the usual order, with
diagonal $T_3$ and $T_8$. 

This indexing system and the explicit expression in terms of commutators of the generators (normalized as $Tr(T^aT^b)=\delta^{ab}/2$)
\begin{equation}
    f^{abc}=-2i{\rm Tr} \left[[T^a,T^b],T^c\right]\;,
\end{equation}
that makes their total antisymmetry explicit, allows to find directly programmable expressions reported by Bossion and Huo~\cite{Bossion:2021zjn},
\begin{eqnarray}    
    f_{S_{nm}S_{kn}A_{km}}&=f_{S_{nm}S_{nk}A_{km}}\nonumber \\=f_{S_{nm}S_{km}A_{kn}}&=f_{A_{nm}A_{kn}A_{km}}=\frac{1}{2} \nonumber \\
    f_{S_{nm}A_{nm}D_{m}}&=-\sqrt{\frac{m-1}{2m}} \nonumber \\
    f_{S_{nm}A_{nm}D_{n}}&=\sqrt{\frac{n}{2(n-1)}}\nonumber \\
    f_{S_{nm}A_{nm}D_{k}}&=\sqrt{\frac{1}{2k(k-1)}} \hspace{0.5cm} m<k<n \nonumber \\
\end{eqnarray}
(other combinations of the different symmetries are null unless obtained by permutation and antisymmetry, for example if $f_{123}=1/2$ then $f_{231}=f_{312}=-f_{321}=-f_{132}=-f_{213}=1/2$).

\subsection*{Sums of squared structure constants for SU(3),
partitioning it in 3-gluon and 5-gluon subsets}
Here we list, as an example of the combinations of $\sum f^2$
with various indices that appear in Eq.~(\ref{eq:sistemapp}),
an exhaustive list of these combinations for the example 
in which the eight gluons of $SU(3)$ split into three light ones
(that, in certain cases but not necessarily, can generate an $SU(2)$
subgroup) and five heavier ones. 
The number of combinations of the eight gluons taken three at a time
is  
$$
\begin{pmatrix} 8 \\ 3  \end{pmatrix} = 56\;,
$$
and we list them explicitly in table~\ref{tab:fconstants1} and following.

\begin{table}[h]
	\caption{Sums of squared structure constants necessary for Eq.~(\ref{eq:sistemapp}). 
The rows alternate in shade. The grey shaded ones indicate the $SU(3)$ gluon combination, with
the first three gluons corresponding to the light ones with colour index $d$ and the following
five ones to the heavy ones with index $D$. 
The row with white background immediately below lists, in the first three columns, the corresponding
 $\sum_{\alpha,B}\left( f^{\alpha Bd}\right) ^2$ to each $d$ entry from the row above. 
The remaining columns give $\sum_{\alpha,b}\left( f^{\alpha bD}\right) ^2$, also associated 
to the index $D$ immediately above each entry.}
	\label{tab:fconstants1}%
	\centering
	\begin{tabular}{c|c|c||c|c|c|c|c}
		\rowcolor[rgb]{ .749,  .749,  .749} 1     & 2     & 3     & 4     & 5     & 6     & 7     & 8 \\
		1.00  & 1.00  & 1.00  & 0.75  & 0.75  & 0.75  & 0.75  & 0.00 \\
		\rowcolor[rgb]{ .749,  .749,  .749} 1     & 2     & 4     & 3     & 5     & 6     & 7     & 8 \\
		1.75  & 1.75  & 2.50  & 2.25  & 1.50  & 0.75  & 0.75  & 0.75 \\
		\rowcolor[rgb]{ .749,  .749,  .749} 1     & 2     & 5     & 3     & 4     & 6     & 7     & 8 \\
		1.75  & 1.75  & 2.50  & 2.25  & 1.50  & 0.75  & 0.75  & 0.75 \\
		\rowcolor[rgb]{ .749,  .749,  .749} 1     & 2     & 6     & 3     & 4     & 5     & 7     & 8 \\
		1.75  & 1.75  & 2.50  & 2.25  & 0.75  & 0.75  & 1.50  & 0.75 \\
		\rowcolor[rgb]{ .749,  .749,  .749} 1     & 2     & 7     & 3     & 4     & 5     & 6     & 8 \\
		1.75  & 1.75  & 2.50  & 2.25  & 0.75  & 0.75  & 1.50  & 0.75 \\
		\rowcolor[rgb]{ .749,  .749,  .749} 1     & 2     & 8     & 3     & 4     & 5     & 6     & 7 \\
		2.00  & 2.00  & 3.00  & 2.00  & 1.25  & 1.25  & 1.25  & 1.25 \\
		\rowcolor[rgb]{ .749,  .749,  .749} 1     & 3     & 4     & 2     & 5     & 6     & 7     & 8 \\
		1.75  & 1.75  & 2.50  & 2.25  & 1.50  & 0.75  & 0.75  & 0.75 \\
		\rowcolor[rgb]{ .749,  .749,  .749} 1     & 3     & 5     & 2     & 4     & 6     & 7     & 8 \\
		1.75  & 1.75  & 2.50  & 2.25  & 1.50  & 0.75  & 0.75  & 0.75 \\
		\rowcolor[rgb]{ .749,  .749,  .749} 1     & 3     & 6     & 2     & 4     & 5     & 7     & 8 \\
		1.75  & 1.75  & 2.50  & 2.25  & 0.75  & 0.75  & 1.50  & 0.75 \\
		\rowcolor[rgb]{ .749,  .749,  .749} 1     & 3     & 7     & 2     & 4     & 5     & 6     & 8 \\
		1.75  & 1.75  & 2.50  & 2.25  & 0.75  & 0.75  & 1.50  & 0.75 \\
		\rowcolor[rgb]{ .749,  .749,  .749} 1     & 3     & 8     & 2     & 4     & 5     & 6     & 7 \\
		2.00  & 2.00  & 3.00  & 2.00  & 1.25  & 1.25  & 1.25  & 1.25 \\
		\rowcolor[rgb]{ .749,  .749,  .749} 1     & 4     & 5     & 2     & 3     & 6     & 7     & 8 \\
		2.50  & 1.75  & 1.75  & 1.50  & 1.50  & 0.75  & 0.75  & 1.50 \\
		\rowcolor[rgb]{ .749,  .749,  .749} 1     & 4     & 6     & 2     & 3     & 5     & 7     & 8 \\
		2.50  & 2.50  & 2.50  & 1.50  & 1.50  & 1.50  & 1.50  & 1.50 \\
		\rowcolor[rgb]{ .749,  .749,  .749} 1     & 4     & 7     & 2     & 3     & 5     & 6     & 8 \\
		2.50  & 2.50  & 2.50  & 1.50  & 1.50  & 1.50  & 1.50  & 1.50 \\
		\rowcolor[rgb]{ .749,  .749,  .749} 1     & 4     & 8     & 2     & 3     & 5     & 6     & 7 \\
		2.75  & 2.00  & 2.25  & 1.25  & 1.25  & 2.00  & 1.25  & 1.25 \\
		\rowcolor[rgb]{ .749,  .749,  .749} 1     & 5     & 6     & 2     & 3     & 4     & 7     & 8 \\
		2.50  & 2.50  & 2.50  & 1.50  & 1.50  & 1.50  & 1.50  & 1.50 \\
		\rowcolor[rgb]{ .749,  .749,  .749} 1     & 5     & 7     & 2     & 3     & 4     & 6     & 8 \\
		2.50  & 2.50  & 2.50  & 1.50  & 1.50  & 1.50  & 1.50  & 1.50 \\
		\rowcolor[rgb]{ .749,  .749,  .749} 1     & 5     & 8     & 2     & 3     & 4     & 6     & 7 \\
		2.75  & 2.00  & 2.25  & 1.25  & 1.25  & 2.00  & 1.25  & 1.25 \\
		\rowcolor[rgb]{ .749,  .749,  .749} 1     & 6     & 7     & 2     & 3     & 4     & 5     & 8 \\
		2.50  & 1.75  & 1.75  & 1.50  & 1.50  & 0.75  & 0.75  & 1.50 \\
	\end{tabular}%
\end{table}%

\begin{table}[h]
	\caption{Continued from table~\ref{tab:fconstants1}}
	\label{tab:fconstants2}%
	\centering
	\begin{tabular}{c|c|c||c|c|c|c|c}
		\rowcolor[rgb]{ .749,  .749,  .749} 1     & 6     & 8     & 2     & 3     & 4     & 5     & 7 \\
		2.75  & 2.00  & 2.25  & 1.25  & 1.25  & 1.25  & 1.25  & 2.00 \\
		\rowcolor[rgb]{ .749,  .749,  .749} 1     & 7     & 8     & 2     & 3     & 4     & 5     & 6 \\
		2.75  & 2.00  & 2.25  & 1.25  & 1.25  & 1.25  & 1.25  & 2.00 \\
		\rowcolor[rgb]{ .749,  .749,  .749} 2     & 3     & 4     & 1     & 5     & 6     & 7     & 8 \\
		1.75  & 1.75  & 2.50  & 2.25  & 1.50  & 0.75  & 0.75  & 0.75 \\
		\rowcolor[rgb]{ .749,  .749,  .749} 2     & 3     & 5     & 1     & 4     & 6     & 7     & 8 \\
		1.75  & 1.75  & 2.50  & 2.25  & 1.50  & 0.75  & 0.75  & 0.75 \\
		\rowcolor[rgb]{ .749,  .749,  .749} 2     & 3     & 6     & 1     & 4     & 5     & 7     & 8 \\
		1.75  & 1.75  & 2.50  & 2.25  & 0.75  & 0.75  & 1.50  & 0.75 \\
		\rowcolor[rgb]{ .749,  .749,  .749} 2     & 3     & 7     & 1     & 4     & 5     & 6     & 8 \\
		1.75  & 1.75  & 2.50  & 2.25  & 0.75  & 0.75  & 1.50  & 0.75 \\
		\rowcolor[rgb]{ .749,  .749,  .749} 2     & 3     & 8     & 1     & 4     & 5     & 6     & 7 \\
		2.00  & 2.00  & 3.00  & 2.00  & 1.25  & 1.25  & 1.25  & 1.25 \\
		\rowcolor[rgb]{ .749,  .749,  .749} 2     & 4     & 5     & 1     & 3     & 6     & 7     & 8 \\
		2.50  & 1.75  & 1.75  & 1.50  & 1.50  & 0.75  & 0.75  & 1.50 \\
		\rowcolor[rgb]{ .749,  .749,  .749} 2     & 4     & 6     & 1     & 3     & 5     & 7     & 8 \\
		2.50  & 2.50  & 2.50  & 1.50  & 1.50  & 1.50  & 1.50  & 1.50 \\
		\rowcolor[rgb]{ .749,  .749,  .749} 2     & 4     & 7     & 1     & 3     & 5     & 6     & 8 \\
		2.50  & 2.50  & 2.50  & 1.50  & 1.50  & 1.50  & 1.50  & 1.50 \\
		\rowcolor[rgb]{ .749,  .749,  .749} 2     & 4     & 8     & 1     & 3     & 5     & 6     & 7 \\
		2.75  & 2.00  & 2.25  & 1.25  & 1.25  & 2.00  & 1.25  & 1.25 \\
		\rowcolor[rgb]{ .749,  .749,  .749} 2     & 5     & 6     & 1     & 3     & 4     & 7     & 8 \\
		2.50  & 2.50  & 2.50  & 1.50  & 1.50  & 1.50  & 1.50  & 1.50 \\
		\rowcolor[rgb]{ .749,  .749,  .749} 2     & 5     & 7     & 1     & 3     & 4     & 6     & 8 \\
		2.50  & 2.50  & 2.50  & 1.50  & 1.50  & 1.50  & 1.50  & 1.50 \\
		\rowcolor[rgb]{ .749,  .749,  .749} 2     & 5     & 8     & 1     & 3     & 4     & 6     & 7 \\
		2.75  & 2.00  & 2.25  & 1.25  & 1.25  & 2.00  & 1.25  & 1.25 \\
		\rowcolor[rgb]{ .749,  .749,  .749} 2     & 6     & 7     & 1     & 3     & 4     & 5     & 8 \\
		2.50  & 1.75  & 1.75  & 1.50  & 1.50  & 0.75  & 0.75  & 1.50 \\
		\rowcolor[rgb]{ .749,  .749,  .749} 2     & 6     & 8     & 1     & 3     & 4     & 5     & 7 \\
		2.75  & 2.00  & 2.25  & 1.25  & 1.25  & 1.25  & 1.25  & 2.00 \\
		\rowcolor[rgb]{ .749,  .749,  .749} 2     & 7     & 8     & 1     & 3     & 4     & 5     & 6 \\
		2.75  & 2.00  & 2.25  & 1.25  & 1.25  & 1.25  & 1.25  & 2.00 \\
		\rowcolor[rgb]{ .749,  .749,  .749} 3     & 4     & 5     & 1     & 2     & 6     & 7     & 8 \\
		2.50  & 1.75  & 1.75  & 1.50  & 1.50  & 0.75  & 0.75  & 1.50 \\
		\rowcolor[rgb]{ .749,  .749,  .749} 3     & 4     & 6     & 1     & 2     & 5     & 7     & 8 \\
		2.50  & 2.50  & 2.50  & 1.50  & 1.50  & 1.50  & 1.50  & 1.50 \\
		\rowcolor[rgb]{ .749,  .749,  .749} 3     & 4     & 7     & 1     & 2     & 5     & 6     & 8 \\
		2.50  & 2.50  & 2.50  & 1.50  & 1.50  & 1.50  & 1.50  & 1.50 \\
		\rowcolor[rgb]{ .749,  .749,  .749} 3     & 4     & 8     & 1     & 2     & 5     & 6     & 7 \\
		2.75  & 2.00  & 2.25  & 1.25  & 1.25  & 2.00  & 1.25  & 1.25 \\
		\rowcolor[rgb]{ .749,  .749,  .749} 3     & 5     & 6     & 1     & 2     & 4     & 7     & 8 \\
		2.50  & 2.50  & 2.50  & 1.50  & 1.50  & 1.50  & 1.50  & 1.50 \\
	\end{tabular}%
\end{table}%

\begin{table}[h]
	\centering
	\caption{Continued from tables~\ref{tab:fconstants1} and~\ref{tab:fconstants2}.}
	\label{tab:fconstants3}%
	\begin{tabular}{c|c|c||c|c|c|c|c}
		\rowcolor[rgb]{ .749,  .749,  .749} 3     & 5     & 7     & 1     & 2     & 4     & 6     & 8 \\
		2.50  & 2.50  & 2.50  & 1.50  & 1.50  & 1.50  & 1.50  & 1.50 \\
		\rowcolor[rgb]{ .749,  .749,  .749} 3     & 5     & 8     & 1     & 2     & 4     & 6     & 7 \\
		2.75  & 2.00  & 2.25  & 1.25  & 1.25  & 2.00  & 1.25  & 1.25 \\
		\rowcolor[rgb]{ .749,  .749,  .749} 3     & 6     & 7     & 1     & 2     & 4     & 5     & 8 \\
		2.50  & 1.75  & 1.75  & 1.50  & 1.50  & 0.75  & 0.75  & 1.50 \\
		\rowcolor[rgb]{ .749,  .749,  .749} 3     & 6     & 8     & 1     & 2     & 4     & 5     & 7 \\
		2.75  & 2.00  & 2.25  & 1.25  & 1.25  & 1.25  & 1.25  & 2.00 \\
		\rowcolor[rgb]{ .749,  .749,  .749} 3     & 7     & 8     & 1     & 2     & 4     & 5     & 6 \\
		2.75  & 2.00  & 2.25  & 1.25  & 1.25  & 1.25  & 1.25  & 2.00 \\
		\rowcolor[rgb]{ .749,  .749,  .749} 4     & 5     & 6     & 1     & 2     & 3     & 7     & 8 \\
		1.75  & 1.75  & 2.50  & 0.75  & 0.75  & 0.75  & 1.50  & 2.25 \\
		\rowcolor[rgb]{ .749,  .749,  .749} 4     & 5     & 7     & 1     & 2     & 3     & 6     & 8 \\
		1.75  & 1.75  & 2.50  & 0.75  & 0.75  & 0.75  & 1.50  & 2.25 \\
		\rowcolor[rgb]{ .749,  .749,  .749} 4     & 5     & 8     & 1     & 2     & 3     & 6     & 7 \\
		1.25  & 1.25  & 1.50  & 0.50  & 0.50  & 0.50  & 1.25  & 1.25 \\
		\rowcolor[rgb]{ .749,  .749,  .749} 4     & 6     & 7     & 1     & 2     & 3     & 5     & 8 \\
		2.50  & 1.75  & 1.75  & 0.75  & 0.75  & 0.75  & 1.50  & 2.25 \\
		\rowcolor[rgb]{ .749,  .749,  .749} 4     & 6     & 8     & 1     & 2     & 3     & 5     & 7 \\
		2.00  & 2.00  & 1.50  & 0.50  & 0.50  & 0.50  & 2.00  & 2.00 \\
		\rowcolor[rgb]{ .749,  .749,  .749} 4     & 7     & 8     & 1     & 2     & 3     & 5     & 6 \\
		2.00  & 2.00  & 1.50  & 0.50  & 0.50  & 0.50  & 2.00  & 2.00 \\
		\rowcolor[rgb]{ .749,  .749,  .749} 5     & 6     & 7     & 1     & 2     & 3     & 4     & 8 \\
		2.50  & 1.75  & 1.75  & 0.75  & 0.75  & 0.75  & 1.50  & 2.25 \\
		\rowcolor[rgb]{ .749,  .749,  .749} 5     & 6     & 8     & 1     & 2     & 3     & 4     & 7 \\
		2.00  & 2.00  & 1.50  & 0.50  & 0.50  & 0.50  & 2.00  & 2.00 \\
		\rowcolor[rgb]{ .749,  .749,  .749} 5     & 7     & 8     & 1     & 2     & 3     & 4     & 6 \\
		2.00  & 2.00  & 1.50  & 0.50  & 0.50  & 0.50  & 2.00  & 2.00 \\
		\rowcolor[rgb]{ .749,  .749,  .749} 6     & 7     & 8     & 1     & 2     & 3     & 4     & 5 \\
		1.25  & 1.25  & 1.50  & 0.50  & 0.50  & 0.50  & 1.25  & 1.25 \\
	\end{tabular}%
\end{table}%

As can be seen, the off-diagonal combinations do not all vanish simultaneously, 
meaning that the Lie algebra has no ideals of either dimension 3 nor 5 (we of course know that 
the $\mathfrak{su}(3)$ Lie algebra has no ideal of any dimension, but it is reassuring to see this appear in the tabulated data. One might entertain the hope that a clever way of splitting the structure constants could bring about a breaking of the global symmetry even for a simple Lie algebra, perhaps of large dimension, but we have not found an example yet, 
nor do we know of a theorem (such as the no--go theorem of Vafa and Witten in the fermion sector) that forbids it at this point.

\clearpage

\section*{Acknowledgments}
The authors thank early conversations with Lucas Barbero.
Work partially supported by the EU under grant 824093 (STRONG2020); 
spanish MICINN under  PID2019-108655GB-I00/AEI/10.13039/501100011033,\\ PID2019-106080GB-C21; 
Univ. Complutense de Madrid under research group 910309 and the IPARCOS institute.

This preprint has been issued with number IPARCOS-UCM-23-070



\begin{thebibliography}{}


\bibitem{Cornwall:1981zr}
J.~M.~Cornwall,
Phys. Rev. D \textbf{26} (1982), 1453
doi:10.1103/PhysRevD.26.1453


\bibitem{Cornwall:2015lna}
J.~M.~Cornwall,
Phys. Rev. D \textbf{93} (2016) no.2, 025021
doi:10.1103/PhysRevD.93.025021


\bibitem{Horak:2022aqx}
J.~Horak, F.~Ihssen, J.~Papavassiliou, J.~M.~Pawlowski, A.~Weber and C.~Wetterich,
SciPost Phys. \textbf{13} (2022) no.2, 042
doi:10.21468/SciPostPhys.13.2.042


\bibitem{Alkofer:2003jr}
R.~Alkofer, C.~S.~Fischer, H.~Reinhardt and L.~von Smekal,
Phys. Rev. D \textbf{68} (2003), 045003
doi:10.1103/PhysRevD.68.045003


\bibitem{Fischer:2004ym}
C.~S.~Fischer, F.~J.~Llanes-Estrada and R.~Alkofer,
Nucl. Phys. B Proc. Suppl. \textbf{141} (2005), 128-133
doi:10.1016/j.nuclphysbps.2004.12.020


\bibitem{Buisseret:2009yv}
F.~Buisseret, V.~Mathieu and C.~Semay,
Phys. Rev. D \textbf{80} (2009), 074021
doi:10.1103/PhysRevD.80.074021


\bibitem{Vento:2004xx}
V.~Vento,
Phys. Rev. D \textbf{73} (2006), 054006
doi:10.1103/PhysRevD.73.054006


\bibitem{Bugg:2000zy}
D.~V.~Bugg, M.~J.~Peardon and B.~S.~Zou,
Phys. Lett. B \textbf{486} (2000), 49-53
doi:10.1016/S0370-2693(00)00752-8

\bibitem{Chen:2005mg}
Y.~Chen \textit{et al.}
Phys. Rev. D \textbf{73} (2006), 014516
doi:10.1103/PhysRevD.73.014516




\bibitem{Lee:2016wiy}
C.~H.~Lee and R.~N.~Mohapatra,
JHEP \textbf{02} (2017), 080
doi:10.1007/JHEP02(2017)080


\bibitem{Mohapatra:1979nn}
R.~N.~Mohapatra and B.~Sakita,
Phys. Rev. D \textbf{21} (1980), 1062
doi:10.1103/PhysRevD.21.1062


\bibitem{Dobson:2022crf}
E.~Dobson, A.~Maas and B.~Riederer,
PoS \textbf{LATTICE2022} (2022), 210
doi:10.22323/1.430.0210
[arXiv:2211.16937 [hep-lat]].

\bibitem{Dobson:2022ngz}
E.~Dobson, A.~Maas and B.~Riederer,
Annals Phys. \textbf{457} (2023), 169404
doi:10.1016/j.aop.2023.169404
[arXiv:2211.05812 [hep-lat]].


\bibitem{Vafa:1983tf}
C.~Vafa and E.~Witten,
Nucl. Phys. B \textbf{234} (1984), 173-188
doi:10.1016/0550-3213(84)90230-X



\bibitem{GarciaFernandez:2015jmn}
G.~Garc\'\i{}a Fern\'andez, J.~Guerrero Rojas and F.~J.~Llanes-Estrada,
Nucl. Phys. B \textbf{915} (2017), 262-284
[erratum: Nucl. Phys. B \textbf{949} (2019), 114755]
doi:10.1016/j.nuclphysb.2016.12.010

\bibitem{Llanes-Estrada:2018azk}
F.~J.~Llanes-Estrada and A.~Salas-Bern\'ardez,
Commun. Theor. Phys. \textbf{71} (2019) no.4, 410-416
doi:10.1088/0253-6102/71/4/410



\bibitem{Cucchieri:2011ig}
A.~Cucchieri, D.~Dudal, T.~Mendes and N.~Vandersickel,
Phys. Rev. D \textbf{85} (2012), 094513
doi:10.1103/PhysRevD.85.094513




\bibitem{Li:2019hyv}
S.~W.~Li, P.~Lowdon, O.~Oliveira and P.~J.~Silva,
Phys. Lett. B \textbf{803} (2020), 135329
doi:10.1016/j.physletb.2020.135329



\bibitem{Reinhardt:2018dhg}
H.~Reinhardt, D.~Campagnari and M.~Quandt,
Universe \textbf{5} (2019) no.2, 40
doi:10.3390/universe5020040


\bibitem{Zwanziger:2002sh}
D.~Zwanziger,
Phys. Rev. Lett. \textbf{90} (2003), 102001
doi:10.1103/PhysRevLett.90.102001




\bibitem{Szczepaniak:1995cw}
A.~Szczepaniak, E.~S.~Swanson, C.~R.~Ji and S.~R.~Cotanch,
Phys. Rev. Lett. \textbf{76} (1996), 2011-2014
doi:10.1103/PhysRevLett.76.2011


\bibitem{Christ:1980ku}
N.~H.~Christ and T.~D.~Lee,
Phys. Rev. D \textbf{22} (1980), 939
doi:10.1103/PhysRevD.22.939


\bibitem{Schwinger:1962wd}
J.~S.~Schwinger,
Phys. Rev. \textbf{127} (1962), 324-330
doi:10.1103/PhysRev.127.324


\bibitem{Llanes-Estrada:2000cdq}
F.~J.~Llanes-Estrada and S.~R.~Cotanch,
Phys. Lett. B \textbf{504} (2001), 15-20
doi:10.1016/S0370-2693(01)00290-8


\bibitem{Watson:2013ghq}
P.~Watson and H.~Reinhardt,
Phys. Rev. D \textbf{89} (2014) no.4, 045008
doi:10.1103/PhysRevD.89.045008


\bibitem{Swift:1983fz}
A.~R.~Swift and J.~L.~Rodriguez Marrero,
Phys. Rev. D \textbf{29} (1984), 1823
doi:10.1103/PhysRevD.29.1823


\bibitem{Adler:1984ri}
S.~L.~Adler and A.~C.~Davis,
Nucl. Phys. B \textbf{244} (1984), 469
doi:10.1016/0550-3213(84)90324-9



\bibitem{Llanes-Estrada:2000ozq}
F.~J.~Llanes-Estrada, S.~R.~Cotanch, P.~J.~de A.~Bicudo, J.~E.~F.~T.~Ribeiro and A.~P.~Szczepaniak,
Nucl. Phys. A \textbf{710} (2002), 45-54
doi:10.1016/S0375-9474(02)01090-4


\bibitem{Bertlmann_2008}
R. Bertlmann and P. Krammer, J. Math. Phys. A {\bf 41} (2008) 235303.


\bibitem{Bossion:2021zjn}
D.~Bossion and P.~Huo,
[arXiv:2108.07219 [math-ph]].
\end{thebibliography}
\end{document}